\documentclass[aps,reprint,pre,groupedaddress,longbibliography]{revtex4-2}
\usepackage{graphicx}
\usepackage{times}
\usepackage{xcolor}
\usepackage{amsmath,amssymb,mathrsfs,amsthm}
\usepackage{bbm}
\usepackage[colorlinks=true,allcolors=blue]{hyperref}
\usepackage{bm}
\usepackage{mathtools}
\usepackage{CJK}

\begin{document}
\begin{CJK*}{UTF8}{gbsn}

\title{Matrix approach to generalized ensemble theory for nonequilibrium  discrete systems}

\author{Shaohua Guan (管绍华)}
\email[]{guanphy@163.com}
\affiliation{Defense Innovation Institute, Chinese Academy of Military Science, Beijing 100071, China}
\affiliation{Intelligent Game and Decision Laboratory, Chinese Academy of Military Science, Beijing 100071, China}

\begin{abstract}
A universal and rigorous ensemble framework for nonequilibrium system remains lacking. Here, we provide a concise framework for the generalized ensemble theory of nonequilibrium discrete systems using matrix-based approach. By introducing an observation matrix, we show that any discrete probability distribution can be formulated as a generalized Boltzmann distribution, with observables and their conjugate variables serving as basis vectors and coordinates in a vector space. Within this framework, we identify the minimal sufficient statistics required to infer the Boltzmann distribution. The nonequilibrium thermodynamic relations and fluctuation-dissipation relations naturally emerge from this framework. Our findings provide a new approach to developing generalized ensemble theory for nonequilibrium discrete systems.
\end{abstract}

\maketitle
\end{CJK*}

\section{Introduction}
Efforts to develop ensemble theories for nonequilibrium complex systems aim to generalize the Boltzmann distribution beyond Gibbs ensembles \cite{gibbs1902elementary}. Various generalized ensemble theories have been proposed to handle specific classes of systems, such as Hill's nanothermodynamics for small systems \cite{hill1962thermodynamics,doi:10.1021/nl010027w,bedeaux2023nanothermodynamics}, Edwards' volume ensemble for granular matter \cite{MEHTA19891091,EDWARDS2005114,baule2018edwards}, and the generalized Gibbs ensemble for integrable systems \cite{rigol2007relaxation,caux2012constructing,langen2015experimental}. However, these frameworks remain largely model-specific, and there is still no unifying formalism that systematically encompasses these diverse ensemble constructions under a common structure. Moreover, classical assumptions of equilibrium ensemble theory, including ergodicity, detailed balance, and thermal equilibrium, often break down or become ill-defined in non-physical systems (e.g., social systems or neural networks) as well as in nonequilibrium systems \cite{meyer1998statistical,han2014energy,dettmer2016network,gnesotto2018broken}. Whether ensemble theory can rigorously describe the probability distributions of such systems remains an open question.

Constructing a generalized ensemble framework for complex systems is crucial for understanding nonequilibrium thermodynamics and response behaviors. Approaches like maximum entropy inference \cite{jaynes1957information} construct Boltzmann distributions by maximizing entropy under constraints on observable averages, such as energy or growth rate in biological systems \cite{de2018statistical,guan2024universal}, while large deviation theory \cite{touchette2009large} provides a probabilistic perspective, where thermodynamic potentials like entropy and free energy emerge as rate function and scaled cumulant generating functions \cite{smith2011large,qian2024internal}. Despite this progress, extending these frameworks to nonequilibrium systems faces significant challenges: constraints are often empirical with ambiguous observable-selection criteria, and determining rate functions for complex systems remains inherently difficult \cite{hanel2014multiplicity,saha2016entropy}. Some studies have attempted to map nonequilibrium steady states onto equilibrium behavior by introducing an effective temperature \cite{hayashi2007temperature,puckett2013equilibrating,lippiello2014nonequilibrium,sorkin2024second}. Nevertheless, a unified theoretical framework that rigorously justifies and systematically establishes such correspondences is still missing. Crucially, a general ensemble formalism for nonequilibrium systems, directly analogous to the Gibbs ensemble in equilibrium statistical mechanics, has not yet been developed.

In this work, We address these gaps by developing a unified matrix formalism for ensemble theory in nonequilibrium discrete systems. We demonstrate that every discrete probability distribution admits a generalized Boltzmann representation---a rigorous algebraic structure within a vector space. Observables such as energy and particle number serve as basis vectors in the vector space, with their dual variables-temperature and chemical potential-acting as coordinates. We show that this linear representation is the core of ensemble theory, offering a mathematically exact formalism that transcends the variational scope of maximum entropy principle. This structure enables a unified formalism for describing diverse ensemble theories. Importantly, our framework relies only on the minimal assumption of the existence of a well-defined discrete probability distribution. This makes it naturally applicable to nonequilibrium discrete systems and non-physical systems, where traditional assumptions---such as ergodicity, thermal equilibrium, or physical replicas---do not hold. The conjugate parameters, constructed via matrix transformations, function as tunable variables analogous to effective temperatures. This framework thereby naturally derives nonequilibrium thermodynamic relations and fluctuation-dissipation relations (FDRs). It thus establishes a new framework for probing the thermodynamics and response behavior of systems far from equilibrium.

\section{Matrix representation of discrete probability distribution}
For a discrete system with a probability distribution, the set of microstates is denoted as $\{\sigma_1,\sigma_2,...,\sigma_N\}$. The probability distribution of microstates is represented by the vector $\bm P = (p_1,p_2,\dots,p_N)^\top$. Each microstate has several observables, and the $i$-th observable is denoted by the observable vector $\bm a_i = (a_{i1},a_{i2},\dots,a_{iN})$, where $a_{ij}$ is the $i$-th observable for $\sigma_j$. Suppose that the system has $N$ linearly independent observation vectors, which can be assembled into a full-rank square matrix $\mathbb {A}$. This observation matrix is thus represented as $\mathbb {A} = (\bm a_1,\bm a_2,\dots,\bm a_N)^\top$. Therefore, the product of $\mathbb A$ and the probability vector $\bm P$ yields the vector of observed averages $\bm O = (o_1,o_2,\dots,o_N)^\top$, expressed as
\begin{equation} \label{eq1}
\mathbb A\bm P = \bm O,
\end{equation}
where $o_i$ denotes the average value of the $i$-th observable. To impose the normalization condition \( \sum_{i=1}^N p_i = 1 \), a unit observation vector ($a_{1j}=1$ for all microstates) is assigned in the first row of \( \mathbb{A} \), yielding \( o_1 = 1 \). Hence, $\mathbb A$ is an $N$-dimensional full-rank matrix with fixed $\bm a_1$. $\bm P$ can be uniquely determined from the vector of observed averages $\bm O$, which is
\begin{equation} \label{inverse1}
\bm P=\mathbb A^{-1}\bm O.
\end{equation}

By taking the negative natural logarithm of each component of $\bm P$, one obtains $\bm{I} = -\ln \bm P$, which corresponds to the self-information vector in information theory \cite{cover1999elements}. By multiplying $\bm{I}$ from the left by $({\mathbb A^\top})^{-1}$, one obtains a vector $\bm B= (\mathbb A^\top)^{-1} \bm{I}$ with entries $(b_1,b_2,\dots,b_N)^\top$. Then, the self-information vector can be expressed as
\begin{equation}\label{eq2}
    \bm{I} = -\ln \bm P = \mathbb A^\top \bm B.
\end{equation}
Therefore, the probability of microstate $\sigma_j$ is
\begin{equation}
p_j = \exp\left(-\sum^N_{i=2} b_ia_{ij}\right)/\exp(b_1)
\end{equation}
due to $a_{1j}=1$. This defines a generalized Boltzmann distribution, with the normalization factor $\exp (b_1)$ identified as the partition function $\mathcal{Z}$. For $i>1$, each row vector $\bm{a}_i$ represents a physically measurable observable, and the corresponding coefficient $b_i$ is its thermodynamic conjugate variable.  For convenience, we refer to the vector $\bm B$ as the Boltzmann vector. This vector carries clear physical meaning: its first component, associated with $\ln \mathcal{Z}$, corresponds to a generalized free energy, while each of the remaining components represents a conjugate variable linked to a specific observable---such as the inverse temperature for energy, the chemical potential for particle number, or external magnetic field for spin magnetization. These quantities serve as control parameters that can be tuned to regulate the system.

Eq.~\eqref{eq2} shows that an arbitrary discrete probability distribution can be expressed as a generalized Boltzmann distribution through the observation matrix $\mathbb A$. It demonstrates that the Boltzmann distribution is a specific representation of probability distributions and is not exclusively confined to equilibrium systems.

Eq.~\eqref{eq2} can be expressed as $\bm{I}=\sum_{i=1}^N b_i\bm {a_i}^\top$, where the vector set $\{\bm{a_i}\}_{i=1}^N$ form a basis for an $N$-dimensional vector space, with $\bm{B} = (b_1,b_2,\dots,b_N)^\top$ acting as coordinates. Since $\bm{a_1}$ is fixed, we define the complementary subspace of $\text{span}(\bm{a_1})$ as $\mathcal{V} = \text{span}(\{\bm{a_i}\}_{i=2}^N)$. In $\mathcal{V}$, the vectors $\{\bm{a_i}\}_{i=2}^N$ constitute a complete basis, and $\{b_i\}_{i=2}^N$ represent coordinates within this subspace. The parameter $b_1$ is not independent but is fixed by the normalization condition, given by $\exp(b_1)=\sum_{j=1}^N \exp(-\sum_{i=2}^Nb_i a_{ij})$. Consequently, $\bm{I}$ is uniquely determined by the basis vectors and coordinates within $\mathcal{V}$. This vector space representation demonstrates that distinct choices of basis vectors in $\mathcal{V}$ lead to different coordinate representations $\{b_i\}_{i=2}^N$ for a given probability distribution. Crucially, the infinite degrees of freedom in selecting basis vectors for $\mathcal{V}$ imply that a probability distribution admits infinitely many equivalent Boltzmann distribution forms. The boundaries of vector spaces are discussed in Appendix \ref{A1}.

Under a change of basis in $\mathcal{V}$, $\mathbb{A}$ is transformed to $\mathbb T\mathbb A$, where $\mathbb T$ is an invertible $N\times N$ matrix. Eq.~\eqref{eq1} becomes $ \mathbb T\mathbb A\bm P=\mathbb T\bm O$, indicating that both $\mathbb{A}$ and $\bm O$ transform by being left-multiplied by $\mathbb T$. To preserve normalization (i.e., the first row of $\mathbb T\mathbb A$ consists of all ones), $\mathbb T$ must satisfy the constraint $T_{1j}=\delta_{1j}$ for all microstates. The generalized Boltzmann distribution then takes the form $-\ln \bm P = \mathbb A^\top \mathbb T^\top (\mathbb T^\top)^{-1}\bm B$, where the Boltzmann vector transforms as $\bm{B}\to (\mathbb T^\top)^{-1}\bm B$. This means that when the observations change, the probability distribution remains unchanged, leading to gauge freedom in statistical mechanics. See Appendix \ref{GF} for details.

\section{Spin Models as an Illustrative Example}
Although any full-rank matrix with the fixed first row $\bm{a_1}$ can mathematically be used to represent a generalized Boltzmann distribution, the observation matrix must be selected to ensure that it is physically interpretable and measurable. Spin models are widely used in combinatorial optimization \cite{mezard1987spin,lucas2014ising}, neural networks \cite{amit1985spin,salakhutdinov2009deep,fischer2012introduction}, and the modeling of biological \cite{agliari2017retrieving} and social systems \cite{lee2015statistical}. Despite their nonequilibrium or non-Hamiltonian in nature, these systems are often described using equilibrium distributions of spin models. Why are such distributions effective? We address this question by analyzing the matrix representation of spin-model probabilities.

\begin{figure}[htbp]
    \centering
    \includegraphics[width=1\linewidth]{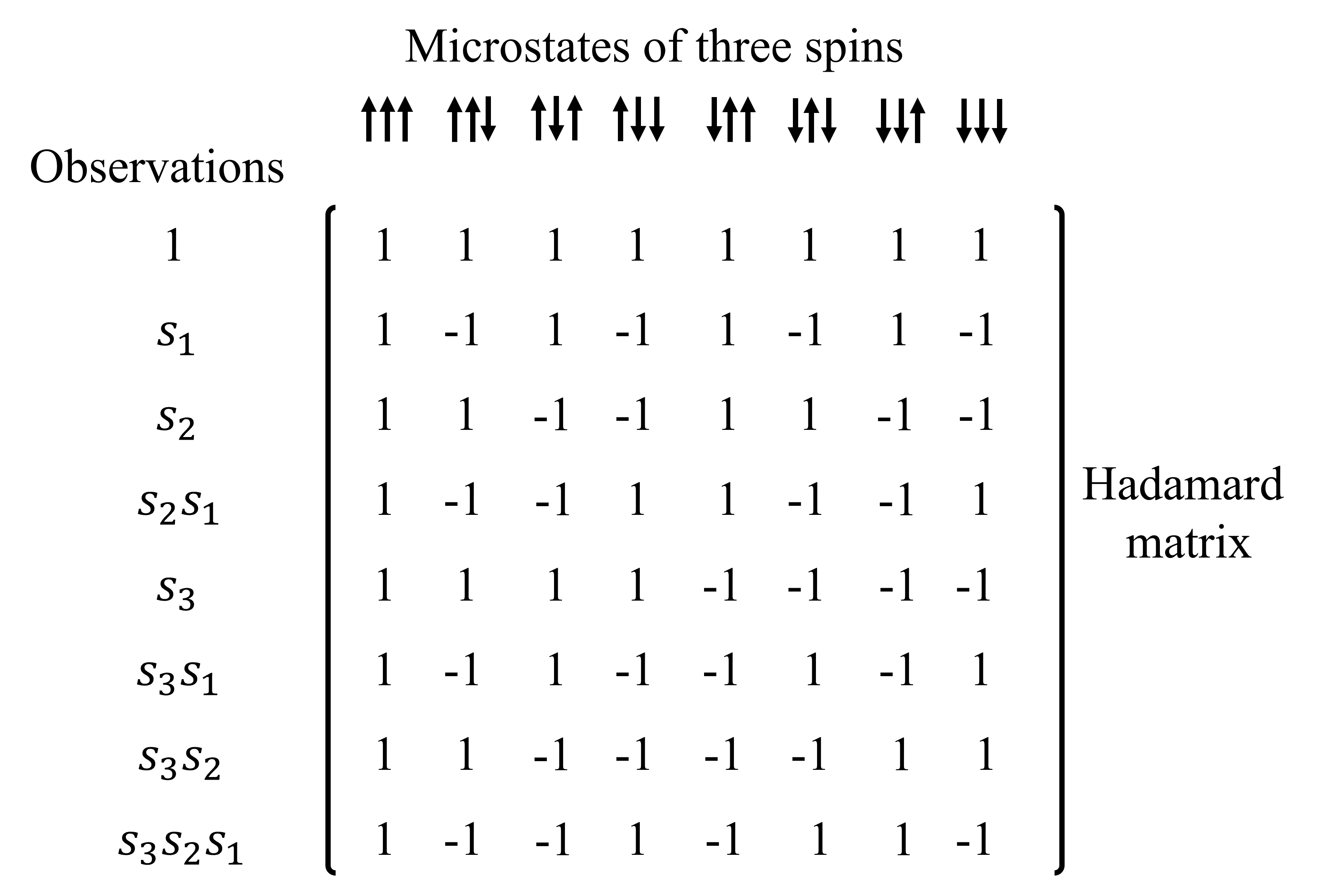}  
    \caption{The illustration of matrix representation of 3-spin model with binary state ($\pm1$). Spin microstates are ordered from spin~3 to spin~1. The left column shows observables evaluated for each microstate, such as $s_2 s_1$, which denotes the product of spins~2 and~1. Each configuration includes a complete set of observables ranging from single-spin terms to higher-order products, ultimately forming the Hadamard matrix. The first row is the normalized vector.}
    \label{fig1}
\end{figure}

The Sylvester Hadamard matrix $\mathbb{H}$ \cite{horadam2012hadamard} provides a natural observation matrix for spin models with $n$ binary variables (shown in Fig.~\ref{fig1}). Its first row consists entirely of ones, enforcing normalization, while the remaining rows represent spin products, ranging from single-spin observables to full $n$-spin products (see Appendix \ref{SH} for details). The Boltzmann distribution based on the Hadamard matrix takes the form $-\ln \bm{P} = \mathbb{H}^\top \bm B$. For a given state $\sigma_j$, the probability is
\begin{equation}
p_j =\frac{1}{\exp{(b_1)}} \exp{(-\sum_{i=2}^{2^n} b_i h_{ij})},\label{ising}
\end{equation}
where $h_{ij}$ denotes the ($i,j$)-th entry of $\mathbb{H}$. This expression can be compared to the Boltzmann distribution of a spin system at equilibrium. The term $\exp(b_1)$ plays the role of the partition function, while $h_{ij}$ corresponds to the measurement of the $i$-th observable on microstate $\sigma_j$, which can represent single-spin quantities, pairwise spin products, and up to $N$-spin products. For $i>1$, the conjugate variable $b_i$ represents the dimensionless interaction strength, defined by $b_i = J_i/k_B T$. Interactions $J_i$ associated with single-spin observables correspond to external fields, while those associated with multi-spin products represent multi-body interactions. 

As shown in Eq.~\eqref{ising}, the probability distribution of any $n$-binary system can be expressed in the form of a generalized Boltzmann distribution. This distribution structure directly corresponds to an equilibrium spin model with multi-order interactions. This explains why such a representation remains effective even for nonequilibrium and non-Hamiltonian systems. In practice, second-order spin models are often employed to approximate higher-order spin interactions when modeling complex systems \cite{schneidman2006weak,tkavcik2013simplest}. It is important to note, however, that although these nonequilibrium systems are modeled using equilibrium-like spin models, the inferred interactions effectively encode nonequilibrium characteristics rather than equilibrium properties.

\section{Conservation laws, symmetry and ergodicity breaking reduce the rank of $\mathbb{A}$}
Physical systems are often constrained by symmetries and conservation laws, which restrict the probability distribution to a lower-dimensional manifold. These features may introduce linear dependencies within the observation matrix $\mathbb{A}$, thereby reducing its effective rank. For conservation laws, if an observable remains constant across all microstates, its corresponding row vector $\bm{a_{cons}}$ in $\mathbb{A}$ becomes linearly dependent on the first row $\bm{a_1}$, which encodes the normalization condition. Including such a row reduces the rank of $\mathbb{A}$. Likewise, symmetries such as rotations or reflections may render distinct microstates observationally indistinguishable, resulting in identical columns in $\mathbb{A}$ and further rank reduction.  An illustrative example of dimensional reduction is shown in Appendix \ref{MR}.

Breakdown of ergodicity can also lead to a reduction in the effective dimensionality of $\mathbb{A}$. When certain microstates are dynamically inaccessible, the corresponding probabilities vanish, resulting in divergent self-information $-\ln p_i \to \infty$. To ensure that the generalized Boltzmann representation remains well-defined, the analysis must be restricted to the accessible subspace, which allows us to remove the columns of $\mathbb{A}$ associated with zero-probability states. This process induces linear dependencies among the remaining observables and thus reduces the rank of $\mathbb{A}$. A minimal full-rank representation could then be constructed to reflect the true support of the distribution.

By reducing a rank-deficient matrix to a minimal full-rank representation, one obtains a more compact description of the system. While the original high-dimensional full-rank matrix remains mathematically valid, the reduced representation is often more efficient and physically meaningful, especially in systems with strong symmetries or conservation constraints.

\section{Observation and Inference}
\begin{figure}[htbp]
    \centering
    \includegraphics[width=0.8\linewidth]{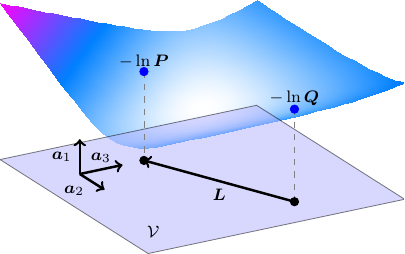} 
    \caption{For a system with three microstates $\{ \sigma_1, \sigma_2, \sigma_3 \}$, assuming that $\bm{a_1}$ is orthogonal to the subspace $\mathcal{V}=\text{span}\{\bm{a_2},\bm{a_3}\}$. $b_2$ and $b_3$ serve as coordinates on this plane, while $b_1$ is determined as a function of $b_2$ and $b_3$, forming a curved surface. The reference distribution $\bm{Q}$ and the target distribution $\bm{P}$, as labeled on the surface, can be projected onto the plane $\mathcal{V}$. The vector $\bm{L}$, defined as the difference between mapped points, can be expressed in terms of $\bm{a_2}$ and $\bm{a_3}$.}
    \label{fig2}
\end{figure}

The inverse problem of determining the Boltzmann vector $\bm{B}$ from the vector of observed averages $\bm{O}$  requires solving
\begin{equation}\label{inf}
\bm{B} = -{(\mathbb{A}^\top)}^{-1} \ln(\mathbb{A}^{-1}\bm{O}).
\end{equation}
However, in practice, the enormous number of microstates makes it intractable to directly measure all averages and perform matrix computations. Instead, one can assume a known reference distribution $\bm{Q}$ with its corresponding Boltzmann vector $\bm{B^{Q}}$ under the observation matrix $\mathbb{A}$. Meanwhile, the target distribution $\bm{P}$ has the Boltzmann parameter vector $\bm{B}$. By subtracting their self-information vectors, one obtains the relation
\begin{subequations}
\begin{align}\label{eq3}
    -\ln (\bm{P}/\bm{Q}) &= \mathbb{A}^\top (\bm{B}-\bm{B^{Q}})=\mathbb{A}^\top \bm{B^{KL}}\\
    &= \underbrace{b_1^{KL}\bm{a_1}^\top}_{\text{Normalization term}} + \underbrace{\sum_{i=2}^N b_i^{KL}\bm{a_i}^\top}_{\text{Difference vector } \bm{L}}.
\end{align}
\end{subequations}
The quantity $-\ln(\bm{P}/\bm{Q})$ represents the difference vector between $\bm{P}$ and $\bm{Q}$ in the self-information space, and its negative inner product with $\bm{P}$ gives the Kullback-Leibler (KL) divergence $D_{KL}(\bm{P}||\bm{Q})$. The term $\bm{B^{KL}}$ denotes their relative coordinates, where the first component is given by $b_1^{KL} \coloneqq b_1-b_1^{Q}=\ln(\mathcal{Z}/\mathcal{Z}^Q)$, with $\mathcal{Z}^Q$ being the partition function of $\bm{Q}$. The remaining components ($\{b_i^{KL} \coloneqq  b_i - b_i^Q\}_{i=2}^N$) represent the coordinate displacements in $\mathcal{V}$, resulting in a difference vector $\bm{L} =  \sum_{i=2}^N b_i^{KL}\bm{a_i}^\top$ (shown in Fig.~\ref{fig2}). This decomposition reveals that the difference vector $-\ln(\bm{P}/\bm{Q})$ separates into two terms: the normalization term and the difference vector $\bm{L}$ in $\mathcal{V}$. We note that the right side of Eq.~\eqref{eq3} plays the same role as the action in dynamical ensembles \cite{xing2019action}, as well as the negative entropy production in the detailed fluctuation relation of stochastic thermodynamics \cite{crooks1999entropy,seifert2012stochastic,maes2020frenesy}.

For a given matrix $\mathbb{A}$ and $\bm{Q}$, $\bm{B}$ and $\bm{B^Q}$ may share the same components at specific indices, implying that certain entries of $\bm{B^{KL}}$ vanish. We define a set $\mathcal{K} = \{ i \mid b_i^{KL} \neq 0, \, i \neq 1 \}$ with $k$ elements, which identifies the non-zero components of $\bm{B^{KL}}$  (excluding the normalization term). The difference vector in the subspace $\mathcal{V}$ then becomes $\bm{L}=\sum_{i\in \mathcal{K}} b_i \bm{a_i}^\top$. The Boltzmann distribution thus takes the form
\begin{align}\label{minkl}
\bm{P}&=\frac{\bm{Q} \exp{(-\sum_{i\in \mathcal{K}}b_i^{KL}\bm{a_i}^\top})}{\exp{(b_1^{KL})}},
\end{align}
where $\exp{(b_1^{KL})}$ ensures normalization. This modified Boltzmann distribution is the solution of the minimum KL divergence $D_{KL}(\bm{P}||\bm{Q})$ inference (Abbreviated as minKL inference) \cite{zdeborova2016statistical} under the constraints of observed averages $\{o_i\}_{i\in \mathcal{K}}$ (see Appendix \ref{AKL} for derivation). Knowledge of $\bm{Q}$ allows the Boltzmann representation of $\bm{P}$ to be fully determined by a small set of observed averages $\{o_i\}_{i\in \mathcal{K}}$. Consequently, $\{o_i\}_{i\in \mathcal{K}}$ serve as the minimal sufficient statistics \cite{fisher1922mathematical,halmos1949application} for $\bm{P}$, eliminating redundant observables while preserving all critical information about the distribution. When $\bm{Q}$ is the uniform distribution, $\bm{B^Q}$ vanishes except for the first component. Eq.~\eqref{minkl} simplifies to $p_j= \exp{(-\sum_{i \in \mathcal{K}} b_i a_{ij}})/\mathcal{Z}$. This corresponds to Jaynes' maximum entropy framework: the distribution \( \bm{P} \) maximizes entropy subject to the constraints of observed averages \( \{o_i\}_{i \in \mathcal{K}} \), representing a special case of the minKL inference. Although the reference distribution $\bm{Q}$ is mathematically arbitrary, it is typically chosen to minimize the number of nonzero components in the Boltzmann vector by incorporating known symmetries or constraints. For example, a uniform $\bm{Q}$ naturally applies to highly symmetric systems and recovers classical equilibrium ensembles.

When observations are insufficient to fully determine the distribution, the minKL inference behaves analogously to Bayesian updating. The reference distribution $\bm Q$ plays the role of a prior $\bm {Q_0}$, and the observed average values serve to update this prior. The update process takes the form
\begin{subequations}
\begin{align}
    -\ln \bm{Q_1} &= -\ln \bm {Q_0} + \mathbb{A}\bm{B^{KL_1}},\\
    \bm{B^{Q_1}} &= \bm{B^{Q_0}} + \bm{B^{KL_1}},
\end{align}
\end{subequations}
where $\bm{B^{KL_1}}$ represents the update vector derived from the minKL inference under limited observations.  
As more observables are incorporated, both the reference distribution $\bm Q_0$ and its associated $\bm{B_{Q_0}}$ are progressively updated to $\bm {Q_k}$ and $\bm{B_{Q_k}}$, and the approximation to the true distribution $\bm P$ becomes increasingly accurate. In the limiting case where a complete set of sufficient statistics is available, the distribution is fully recovered.

\section{Embedding Classical Ensembles into a Unified Framework}
In classical ensembles, the reference distribution $\bm Q$ is typically taken to be uniform, and the target distribution $\bm{P}$ is characterized by the difference vector $\bm{L}=\sum_{i\in \mathcal{K}} b_i \bm{a_i}^\top$, where only a subset of observables carry nonzero conjugate parameters. For example, the canonical ensemble corresponds to a single observable---the Hamiltonian---with $b_2 = 1/k_B T$. The grand canonical ensemble includes both the energy $a_{2j}=E(\sigma_j)$ and the particle number observable $a_{3j}=N_{par}(\sigma_j)$, where $b_2 = 1/k_BT$ and $b_3=-\mu/k_BT$. The microcanonical ensemble corresponds to the limit $\bm{P} = \bm{Q}$, with all $b_i=0$ except the normalization term. Hill's nanothermodynamics \cite{hill1962thermodynamics,doi:10.1021/nl010027w,bedeaux2023nanothermodynamics} adds subdivision potential as additional observables to describe surface effects in finite systems. The generalized Gibbs ensemble \cite{rigol2007relaxation,caux2012constructing,langen2015experimental}, relevant for integrable systems, includes all local conserved quantities as sufficient observables. These ensemble theories capture the sufficient statistical observables that govern the probability distribution of the system. However, for systems with long-range interactions or complex constraints, such observables may be insufficient to fully characterize the probabilistic behavior, ultimately leading to the failure of classical ensemble models. Our framework demonstrates that by constructing a sufficient set of observables, a generalized Boltzmann distribution can still capture complex probability distributions. The bijective mapping between $\bm B$ and $\bm I$ ensures this generalization. Identifying an appropriate sufficient set of observables for specific complex systems remains an important direction for future investigation.

The matrix-based formalism also accommodates Tsallis statistics \cite{tsallis1988possible} by replacing the self-information with the \textit{q}-deformed form \cite{tsallis2009nonadditive}, $-\ln_q \bm P = \mathbb{A}^\top \bm B$. The Tsallis distribution arises when $\bm B$ includes a normalized term $\ln_q \mathcal{Z}$ and a nonzero coefficient associated with the vector of Hamiltonian. This matrix representation yields a generalized form of Tsallis distribution, structurally analogous to the generalized Boltzmann form, and shows that both distributions can be embedded within a unified matrix framework. More details are shown in Appendix \ref{ATS}.

\section{Thermodynamic relation and fluctuation-dissipation relations}
The matrix formalism extends the equilibrium ensemble framework to nonequilibrium discrete systems, enabling a unified description of probability distributions that preserves key thermodynamic properties. The thermodynamic relation in nonequilibrium discrete systems is
\begin{align}
        S &= -\bm{P}^\top \ln \bm{P} = \bm{P}^\top \mathbb{A}^\top \bm{B}= \bm{O}^\top \bm{B}\\
    & = \ln \mathcal{Z} + \sum^N_{i=2} b_io_i.
\end{align}
This provides a generalized Legendre-type thermodynamic structure for nonequilibrium discrete systems, in direct analogy with equilibrium ensemble theory. The FDRs connect a system's linear response to its intrinsic fluctuations. Specifically, consider a set of observables whose fluctuations are described by the covariance
\begin{align}
\chi_{ij} = \mathrm{Cov}(\bm{a_i}, \bm{a_j}) = \sum_{k=1}^N p_k\, a_{ik} a_{jk} - o_io_j ,
\end{align}
and whose response is quantified by the susceptibility $\partial o_i/\partial b_j$. Within our framework, the generalized Boltzmann distribution yields the linear relation
\begin{align}
    -\partial o_i/\partial b_j = \chi_{ij} \quad   (i>1,j>1),
\end{align}
which remains valid even when the system operates far from detailed balance.

Several studies characterize nonequilibrium system behavior by introducing an effective temperature or modified FDRs \cite{harada2005equality,baiesi2009fluctuations,prost2009generalized,seifert2010fluctuation,altaner2016fluctuation,aslyamov2025nonequilibrium}. While these works provide valuable case-specific insights, our framework reconstructs state-function-like thermodynamic relations for nonequilibrium discrete systems. This breakthrough enables the FDRs to be expressed in a form that is an exact analogue of equilibrium thermodynamics. It does not merely re-derive such quantities. Instead, it systematically generalizes this paradigm by establishing the mathematical foundation that rigorously defines effective temperatures and guarantees the validity of FDRs across any nonequilibrium discrete systems. This structural unification constitutes the important advance of our work.

\section{Case study: Markov jump processes}
For Markov jump processes, $w_{ij}$ denotes the transition rate from state $\sigma_j$ to $\sigma_i$. We construct the flux matrix $\mathbb{J}$ with the first row uniformly set to unity to enforce normalization. For $i > 1$, the $i$-th row consists of $w_{i,i-1}$ at the $(i-1)$-th position and $-w_{i-1,i}$ at the $i$-th position, with all other entries equal to zero. The observed averages $\bm{O}$ correspond to the net fluxes $\langle J_i \rangle= o_i =  p_{i-1}w_{i,i-1} - p_iw_{i-1,i} \quad (i > 1)$, and the Boltzmann distribution is
\begin{equation}
p_i = \frac{\exp\left(b_i w_{i-1,i} - b_{i+1}w_{i+1,i}\right)}{\mathcal{Z}},
\end{equation}
with boundary conditions $w_{01} = w_{N+1,N} = 0$. 

For equilibrium and nonequilibrium systems sharing identical steady-state probability distributions, static observations of microstates cannot discriminate between these two regimes. A critical distinction lies in the detailed balance condition: in equilibrium, the probability flux between any two microstates satisfies $p_i w_{ji} = p_j w_{ij}$. To unambiguously classify a system's steady state, dynamical details---specifically transition rates $w_{ij}$, which are computable from microstate trajectory data---must be incorporated into the observation matrix. These rates encode non-equilibrium signatures by violating detailed balance conditions, allowing the distinction between equilibrium and non-equilibrium states. Under detailed balance conditions ($p_i w_{ji} = p_j w_{ij}$), all net fluxes vanish, reflecting equilibrium. The bijective mapping between $\bm{P}$ and $\bm{O}$ via the full-rank $\mathbb{J}$ ensures equilibrium distributions yield vanishing net fluxes in $\bm{O}$, while nonequilibrium steady state exhibit nonzero net fluxes.

For Markov jump processes, the thermodynamic relation is
\begin{equation}
    S = \ln \mathcal{Z} + \sum_{i=2}^N b_i \langle J_i \rangle,
\end{equation}
and the FDR is 
\begin{equation}
    -\partial \langle J_i \rangle/ \partial b_j = \langle J_iJ_j \rangle - \langle J_i \rangle\langle J_j \rangle.
\end{equation}

Under the substitution $b_i = 1/t_i$ ($i>1$), it becomes
\begin{equation}
    t_j^2 \partial \langle J_i \rangle/\partial t_j = \langle J_iJ_j \rangle - \langle J_i \rangle\langle J_j \rangle,
\end{equation}
where $t_j$ acts as an \textit{effective temperature} governing edge flux fluctuations. The covariance $\langle J_iJ_j \rangle - \langle J_i \rangle\langle J_j \rangle$ quantifies fluctuations in the edge fluxes, while the left-hand side represents the response to effective temperature variation.

This framework defines effective temperature for each net flux and establishes nonequilibrium thermodynamic relation involving entropy, free energy, and flux-temperature conjugate pairs. It retains the structural form of equilibrium thermodynamics while extending it to systems far from equilibrium, offering both interpretability and analytical tractability. Recent studies \cite{owen2020universal,maes2020frenesy,zheng2025universal,aslyamov2024nonequilibrium,aslyamov2024general,ptaszynski2024nonequilibrium,aslyamov2025nonequilibrium} have concentrated on the response behavior of nonequilibrium Markov processes, establishing several equalities and bounds. Notably, Ref.~\cite{aslyamov2024nonequilibrium,aslyamov2024general,ptaszynski2024nonequilibrium,aslyamov2025nonequilibrium} also employ a matrix approach to study response behavior. In \cite{aslyamov2024nonequilibrium}, They utilize a transition-rate matrix that is structurally similar to the observation matrix but differs only in the placement of the unit vector, and perturb an equation analogous to Eq.~\eqref{eq1}. In contrast, our approach leverages Eq.~\eqref{eq2} to directly establish the link between fluctuations and response, resulting in the fluctuation-dissipation relation. Detailed mathematical connections are derived in Appendix \ref{DNR}.

\section{Discussions}
We have established a generalized ensemble framework that characterizes thermodynamic behavior through matrix representations of probability distributions, requiring neither ergodicity nor equiprobability assumptions. Our key contribution is developing a mathematical formalism for the ensemble theory of discrete systems, which offers a precise algebraic framework for describing any discrete probability distribution in nonequilibrium systems. This extends ensemble theory beyond equilibrium to encompass any system admitting a probabilistic description. This algebraic structure enables the discovery of effective temperature and thermodynamic relations directly from distributions. Our framework goes beyond Janeys' framework, possessing a complete mathematical structure and is capable of rigorously representing nonequilibrium discrete probabilities.

The construction of the observation matrix is a crucial issue. In our work, we propose three types of observation matrices: the Hadamard matrix, the flow matrix, and the Vandermonde matrix (shown in Appendix \ref{Van}). The flux matrix and Hadamard matrix serve distinct purposes, depending on accessible observables. If probability currents are measurable, the flux matrix is appropriate. However, for binary-spin systems where only spin configurations are observed, currents are inaccessible, necessitating the Hadamard basis. Note that the Hadamard representation alone cannot distinguish equilibrium from nonequilibrium states, because it cannot access the system's dynamical information. The nature of the observable data---directly extracted from experiments or simulations (e.g., local energy, spin, density)---determines the choice of observation basis vectors, thereby governing the physical interpretation of the Boltzmann vector $\bm{B}$.

While we illustrated examples such as the Hadamard matrix for spin systems, the flux matrix for Markov processes, and the Vandermonde matrix for random walks (Appendix \ref{Van}), systematically deriving compact, physically informed matrix representations by exploiting symmetries, constraints, or couplings remains a key open challenge.

\begin{acknowledgments}
S.G. gratefully acknowledges helpful discussions with Hualin Shi and Yanting Wang (ITP, UCAS), and with Xiao Han (Beijing Jiaotong University).
\end{acknowledgments}

\appendix
\section{Vector Spaces and Their Boundaries \label{A1}}

For a probability distribution of $N$ microstates, the space it occupies is the $(N-1)$-dimensional probability simplex, denoted as:  
\begin{equation}
    \Delta^{N-1} = \left\{ \bm{P} = (p_1, \dots, p_N)^\top \in \mathrm{R}^N \ \middle|\ p_i \geq 0, \sum_{i=1}^{N} p_i = 1 \right\}.
\end{equation}
This simplex is a convex subset of the \((N-1)\)-dimensional affine hyperplane in \( \mathrm{R}^N \) defined by the normalization constraint $ \sum_{i=1}^{N} p_i = 1 $. The boundary of $ \Delta^{N-1} $ consists of points where at least one coordinate  $p_i$  is zero, forming lower-dimensional subsimplices.

Then, we consider a full-rank linear transformation $\mathbb{A}\bm{P}=\bm O$, where $\mathbb{A} \in \mathrm{R}^{N \times N} $ is an invertible matrix with its first row consisting entirely of ones. Since $\mathbb{A}$ is full-rank, the transformation is bijective, mapping  $\Delta^{N-1}$  onto a new affine subspace of  $\mathbb{R}^N$. Specifically, since the first row of $\mathbb{A}$ sums the components of $\bm P$, the first component of $\bm O$ is always 1. Thus, the space $ \mathcal{A}_O$ containing $\bm O$ is an ($N-1$)-dimensional affine subspace given by:  
\begin{equation}
    \mathcal{A}_O = \left\{ \bm{O} \in \mathrm{R}^N \ \middle|\ \bm{e_1} \bm{O} = 1  \right\},
\end{equation}
where $\bm{e_1} = (1, 0, \dots, 0)$. The constraints on the vector $\bm O$ originate from those on the probability $\bm P$, requiring that each element of $\mathbb{A}^{-1}\bm{O}$ be non-negative.

Taking the natural logarithm of each coordinate in the probability simplex generates the space of self-information $\bm{I}$, which is an ($N-1$)-dimensional manifold in $\mathrm{R}^{N}$ with the sum constraint $\sum_{i=1}^N \exp{(-I_i)} =1$ and boundary constraints $I_i\geq 0$. The space of vector $\bm B$ is a full-rank linear transformation of the self-information space, which is also an ($N-1$)-dimensional manifold in $\mathrm{R}^{N}$ with the sum constraint $\sum_{j=1}^N\exp{(-\sum_{i=1}^Nb_i a_{ij})} =1$ and each element of $\mathbb{A}^\top\bm{B}$ is non-negative.

\section{Gauge Freedom \label{GF}}

Although vectors $\bm O$ and $\bm B$ are influenced by the choice of the observation matrix $\mathbb{A}$, the probability distribution remains invariant under the transformation by matrix $\mathbb T$, which reflects the gauge freedom in statistical mechanics. A simple example is the selection of the zero point for the observable. For a given observation matrix $\mathbb A$, shifting the zero point of the $i$-th observable by $x_0$ is equivalent to modifying the $i$-th row to $\bm a_i + x_0\bm a_1$. This matrix transformation can be achieved by left-multiplying an elementary row transformation matrix $\mathbb T_x$, where the $i$-th row and first column of $\mathbb T_x$ is $x_0$, the diagonal elements are 1, and all other entries are 0. For example, in the case of $N=5$, when the observable $\bm{a_4}$ is shifted to $\bm{a_4}+x_0\bm{a_1}$, the corresponding matrix is   
\begin{equation}
\mathbb{T}_x = 
\begin{bmatrix}
1 & 0 & 0 & 0 & 0 \\
0 & 1 & 0 & 0 & 0 \\
0 & 0 & 1 & 0 & 0 \\
x_0 & 0 & 0 & 1 & 0\\
0 & 0 & 0 &0 & 1 
\end{bmatrix}.
\end{equation}
The Boltzmann vector $\bm B$ is transformed via the matrix $(\mathbb{T}_x^\top)^{-1}$, which is
\begin{equation}
(\mathbb{T}_x^\top)^{-1} = 
\begin{bmatrix}
1 & 0 & 0 & -x_0 & 0 \\
0 & 1 & 0 & 0 & 0 \\
0 & 0 & 1 & 0 & 0 \\
0 & 0 & 0 & 1 & 0\\
0 & 0 & 0 &0 & 1 
\end{bmatrix}.
\end{equation}
Then, only the first entry of the transformed matrix $(\mathbb{T}_x^\top)^{-1}\bm{B}$ is modified to $b_1 - x_0 b_i$. This implies that, after shifting the zero point of a specific observable, only the corresponding observable and the partition function are altered, while all other observables, conjugate variables, and the probability distribution remain unchanged.

\section{Spin model and Hadamard matrix \label{SH}}

For an $n$-spin system with binary states ($\pm 1$), microstates are enumerated through the tensor product construction $\bm{M} = (u_n,d_n) \otimes \cdots \otimes (u_1,d_1)$, where $u_i$ and $d_i$ denote spin up and down. Observables $\bm{S} = (1,s_n) \otimes \cdots \otimes (1,s_1)$ generate $2^n$ distinct measurement operators. The first element is $1$ for all microstates, while the remaining terms describe spin products, ranging from single-spin measurements ($s_i$) to full $n$-spin products ($s_1\cdots s_n$). Each observable operator acts on microstates to produce $\pm 1$ values via spin product evaluations. This structured matrix construction directly yields the $2^n \times 2^n$ Sylvester Hadamard matrix $\mathbb{H}$ \cite{horadam2012hadamard}, where element $h_{ij}$ equals the measurement of the $i$-th observable in $\bm{S}$ applied to the $j$-th microstate in $\bm{M}$.

Sylvester's construction recursively generates Hadamard matrices  $\mathbb{H}_{2^n}$ starting from  
\begin{equation}
\mathbb{H}_1 = \begin{bmatrix} 1 \end{bmatrix},
\end{equation}
and for \( n \geq 1 \),
\begin{equation}
\mathbb{H}_{2^n} = \begin{bmatrix} \mathbb{H}_{2^{n-1}} & \mathbb{H}_{2^{n-1}} \\ \mathbb{H}_{2^{n-1}} & -\mathbb{H}_{2^{n-1}} \end{bmatrix}.
\end{equation}
This yields  $\mathbb{H}_{2^n} = \mathbb{H}_2^{\otimes n}$, where 
\begin{equation}
\mathbb{H}_2 = \begin{bmatrix} 1 & 1 \\ 1 & -1 \end{bmatrix}.
\end{equation}
For spin systems, starting with a single spin, the microstate is represented by $\bm{M_1} = (u_1,d_1)$, and the observables are $\bm{S_1} =  (1,s_1) $. The corresponding observation matrix, based on the order of microstates and observables, is given by
\begin{equation}
\begin{array}{c|cc}
  & u_1 & d_1 \\
\hline
1 & 1 & 1 \\
s_1 & 1 & -1 \\
\end{array}
= \mathbb{H}_2,
\end{equation}
where the first entry of $\bm{S_1} =  (1,s_1) $ represents a value of $1$ for all microstates, and the second entry corresponds to the measurement of the spin $s_1$ on $\bm{M_1} = (u_1,d_1)$, yielding $(1,-1)$. 

When an additional spin is introduced, the microstates become $\bm{M_2} = (u_2,d_2) \otimes  (u_1,d_1)=(u_2u_1,u_2d_1,d_2u_1,d_2d_1)$, and the observables are $\bm{S_2} = (1,s_2) \otimes (1,s_1) = (1,s_1,s_2,s_2s_1)$. The observation matrix, following the order of the microstates and observables, is
\begin{equation}
\begin{array}{c|cccc}
  & u_2u_1 & u_2d_1 & d_2u_1 & d_2d_1 \\
\hline
1 & 1 & 1 & 1 & 1 \\
s_1 & 1 & -1 & 1 & -1 \\
s_2 & 1 & 1 & -1 & -1 \\
s_2s_1 & 1 & -1 & -1 & 1 \\
\end{array}
= \mathbb{H}_2 \otimes \mathbb{H}_2 = \mathbb{H}_{2^2}.
\end{equation}
As more spins are added, this process iterates, yielding  $\bm{M_n} = (u_n,d_n)\otimes \bm{M_{n-1}}=(u_n\bm{M_{n-1}},d_n\bm{M_{n-1}})$ and $\bm{S_n} = (1,s_n)\otimes \bm{S_{n-1}}=(\bm{S_{n-1}},s_n\bm{S_{n-1}})$. Then, the observation matrix becomes
\begin{equation}
\begin{array}{c|cc}
  & u_n\bm{M_{n-1}} & d_1\bm{M_{n-1}} \\
\hline
\bm{S_{n-1}} & \mathbb{H}_{2^{n-1}} & \mathbb{H}_{2^{n-1}} \\
s_n\bm{S_{n-1}} & \mathbb{H}_{2^{n-1}} & -\mathbb{H}_{2^{n-1}} \\
\end{array}
=  \mathbb{H}_{2^n} .
\end{equation}

The Boltzmann distribution based on the Hadamard matrix takes the form
\begin{equation}
-\ln p_j = b_1 + \sum_{i=2}^{2^n} b_i h_{ij},
\end{equation}
where $b_i \equiv J_i/k_B T$ ($i>1$) represents the dimensionless ratio of interaction strength ($J_i$) to thermal energy $k_BT$. The coefficients $b_i$ of single-spin $s_i$ map to external magnetic fields, while multi-spin terms encode $k$-body interactions, enabling the construction of desired spin models through parameter constraints. For example, nonzero $b_i$ for spatially separated spins induces long-range interactions, whereas nonzero $b_i$ for $k$-spin ($k>2$) correlations generates higher-order interactions. This universal structure naturally incorporates classical spin models: the 2D Ising model emerges when restricting $b_i \neq 0$ to nearest-neighbor pairs; the Sherrington-Kirkpatrick model \cite{sherrington1975solvable} is realized through Gaussian-distributed $b_i$ for all two-spin terms; and $k$-spin Ising models \cite{fan2011one} are obtained by selectively activating $k$-body couplings. The completeness of $\mathbb{H}$ (spanning all possible spin correlations) ensures this generality.

\section{Example: Dimensional reduction in a conserved system \label{MR}}

Consider a system of $n$ binary spins $s_i = \pm 1$ with conserved total magnetization  
\[
M_{cost} = \sum_{i=1}^n s_i.
\]
We adopt the Sylvester Hadamard matrix $\mathbb{H} \in \mathbb{R}^{2^n \times 2^n}$ as the initial observation matrix, with columns indexed by spin configurations $\bm{\sigma}_j$ and rows corresponding to all possible spin products (from the constant term to the full $n$-body interaction).

In the Hadamard matrix, we can sum all rows representing individual spins to form a new row. For example, row $\bm{a_2}$, originally representing $s_1$, becomes $\sum_{i=1}^n s_i$---the total magnetization. However, since the total magnetization is fixed at $M_{cost}$ for all microstates, then the row $\bm{a_2}$ should be $(M_{cost},M_{cost},...,M_{cost})$. Consequently, row vectors $\bm{a_2}$ and $\bm{a_1}$ are linearly dependent. This shows that a full $2^n$-dimensional Hadamard matrix $\mathbb{H}$ is not required; we can remove one single-spin observable row from the Hadamard matrix, reducing its rank by 1 and achieving dimensional reduction under the conservation constraint.

\section{The Correspondence Between the Minimum KL Divergence and the 
Modified Boltzmann Distribution \label{AKL}}

Given the problem of minimizing the Kullback-Leibler divergence  $D_{KL}(\bm{P} || \bm{Q})$  subject to the observation constraints $ \{ o_i \}_{i\in \mathcal{K}}$, the objective is to find the distribution $ \bm P $ that minimizes

\begin{align}
D_{KL}(\bm{P} || \bm{Q}) = \sum_j p_j \ln \frac{p_j}{q_j}.
\end{align}

The constraints are
\begin{align}
    \sum_j p_j a_{ij} &= o_i, \quad \forall i\in \mathcal{K},\\
    \sum_j p_j &= 1.
\end{align}
To solve this constrained optimization problem using the method of Lagrange multipliers, we introduce the Lagrange multipliers  $\lambda_i$  for the constraints. The Lagrange function is
\begin{align}
    \mathcal{L}(\bm{P}, \lambda) = &\sum_j p_j \ln \frac{p_j}{q_j} + \sum_i \lambda_i \left( \sum_j p_j a_{ij} - o_i \right)\\
    &+ \gamma \left( \sum_j p_j - 1 \right),
\end{align}
where $ \gamma $ is the Lagrange multiplier for the normalization condition. We take the partial derivative of the Lagrange function with respect to  $p_j$  and set it equal to zero
\begin{align}
    \frac{\partial \mathcal{L}}{\partial p_j} = \ln \frac{p_j}{q_j} + 1 + \sum_i \lambda_i a_{ij} + \gamma = 0.
\end{align}
Thus, the optimal distribution $p_j$ is
\begin{align}
    p_j = q_j \exp \left( - \sum_i \lambda_i a_{ij} - \gamma -1 \right).
\end{align}
Let $\mathcal{Z}_{KL}$ be the partition function
\begin{align}
   \mathcal{Z}_{KL} = \sum_j q_j \exp \left( - \sum_i \lambda_i a_{ij} \right).
\end{align}
Then, we obtain
\begin{align}
    p_j = \frac{q_j}{\mathcal{Z}_{KL}} \exp \left( - \sum_i \lambda_i a_{ij} \right).
\end{align}
This result shows that the optimal distribution $\bm P$ is obtained by reweighting the reference distribution $\bm Q$ using exponential factors that enforce the observation constraints, analogous to the maximum entropy principle in statistical physics \cite{jaynes1957information}.

The solution of the minimum KL divergence inference corresponds to the modified Boltzmann distribution
\begin{align}
\bm{P}&=\frac{\bm{Q} \exp{(-\sum_{i\in \mathcal{K}}b_i^{KL}\bm{a_i}^\top})}{\exp{(b_1^{KL})}}.
\end{align}
The Lagrange multipliers $\lambda_i$ correspond to $b_i^{KL}$, and the partition function $\mathcal{Z}_{KL}$ is equivalent to $\exp (b_1^{KL})$. When the reference distribution $\bm Q$ is chosen as the uniform distribution, the minimum KL divergence inference reduces to the maximum entropy inference.

\section{Matrix representation of Tsallis distribution \label{ATS}}

Tsallis entropy \cite{tsallis1988possible} is given by  
\begin{equation}
    S_q = k \frac{1-\sum_{i=1}^N p_i^q}{q-1} \quad (\sum_{i=1}^N p_i=1).
\end{equation}
We assume $k=1$ and omit it in the following equations. By introducing the $q$-logarithmic and $q$-exponential functions
\begin{align}
    \ln_q x &\equiv \frac{x^{1-q}-1}{1-q} \quad (x>0; \ln_1 x=\ln x),\\
    e_q^x &\equiv [1+(1-q)x]^{\frac{1}{1-q}}_{+} \quad (e_1^x = e^x; [z]_+ \equiv \max(z,0)),
\end{align}
the Tsallis entropy can be written as 
\begin{equation}
    S_q = - \sum_{i=1}^N p_i^q \ln_q p_i.
\end{equation}
The constraint associated with the Tsallis distribution is $U = \sum_{i=1}^N p_i^q \epsilon_i$ (Curado-Tsallis type constraints \cite{Curado_1991}). We note that this constraint provides an equivalent description of Tsallis statistics using the escort average constraint $U = \sum_{i=1}^N p_i^q \epsilon_i/ \sum_{i=1}^N p_i^q$ \cite{ferri2005equivalence}. Under this constraint, the maximum Tsallis entropy yields the Tsallis distribution
\begin{equation}
    p_i =\frac{1}{\mathcal{Z}_q} (1-(1-q)\beta \epsilon_i)^{1/(1-q)} = \frac{1}{\mathcal{Z}_q} e_q^{-\beta \epsilon_i},
\end{equation}
where $\mathcal{Z}_q$ is the partition function of the Tsallis distribution \cite{tsallis2009nonadditive,Tsallis_2010}.

Following the same procedure in the main article, the generalized constraints can be written as 
\begin{equation}
    \mathbb{A}\mathbb{T}_q\bm P = \bm O.
\end{equation}
The matrix $\mathbb{T}_q$ is the transfer matrix which maps each $p_i$ to $p_i^q$ and is defined as $\mathbb{T}_q \equiv \bm {P^q} \bm P^\top (\bm P\bm P^\top)^{-1}$. The $i$-th entry of vector $\bm {P^q}$ is $p_i^q$. Then we have 
\begin{align}
    \mathbb{A}\bm {P^q}& = \bm O,\\
    -\ln_q \bm P &= \mathbb{A}^{\top} \bm B.
\end{align}
When the vector $\bm{B}$ contains only the normalized term $\ln_q \mathcal{Z}$ and the parameter $b_2$ (with all other entries set to zero) and the second row of matrix $\mathbb{A}$ encodes the Hamiltonian, the Tsallis distribution emerges naturally within this framework. The Tsallis entropy is $S_q = - {\bm{P^q}}^\top \ln_q \bm P = {\bm{P^q}}^\top \mathbb{A}^\top \bm B = \bm{O}^\top  \bm B$.   The matrix representation of the Tsallis distribution naturally extends to a generalized Tsallis distribution, analogous to the generalization of the Boltzmann distribution. Therefore, within this unified matrix framework, any probability distribution can, in principle, be represented in either a generalized Boltzmann form or a generalized Tsallis form.

\section{Derivation of nonequilibrium response \label{DNR}}

Since $\mathbb{A} \bm P =  \bm O$, the response of the probability vector with respect to the control parameter $\eta$ is 
\begin{align}\label{re}
   & \partial_{\eta} \mathbb{A}\cdot \bm P + \mathbb{A}\cdot\partial_\eta \bm P = \partial_{\eta}  \bm O,\\
  &   \partial_\eta \bm P = - \mathbb{A}^{-1}\cdot\partial_{\eta} \mathbb{A}\cdot\bm P + \mathbb{A}^{-1} \cdot \partial_{\eta}  \bm O.
\end{align}
A special choice of $\mathbb{A}$ is the matrix $\mathbb{K}_n$ in \cite{aslyamov2024nonequilibrium}. The only difference lies in the position of the unit vector row. This is convenient for computing the response, since the term $\partial_{\eta}  \bm O$ vanishes. By setting $\mathbb{A}$ to be the matrix $\mathbb{K}_n$, the above response equation can be directly reduced to the main result (Eq.~(6)) in \cite{aslyamov2024nonequilibrium}. 

In the main article, we use $-\ln \bm P =  \mathbb{A}^\top \bm B$ to derive a new formula for the response behavior. The response to variations in $b_i$ leads to the fluctuation-dissipation relations. If the control parameter is $\eta$, we can express the change in the observation average $o_i$ with respect to $\eta$ as
\begin{align}
     \partial_\eta o_i &= \sum_{j=2}^N \frac{\partial o_i}{\partial b_j}\frac{\partial b_j}{\partial \eta} =-\sum_{j=2}^N \chi_{ij}\frac{\partial b_j}{\partial \eta}.
\end{align}

\section{Vandermonde matrix \label{Van}}

The Vandermonde matrix $\mathbb V$ is commonly used in polynomial interpolation, where its non-zero determinant ensures the uniqueness of the interpolating polynomial \cite{strang2006linear}. The entries of $\mathbb V$ are defined as $V_{ij}=x_i^{j-1}$ with distinct $x_i$, ensuring that the first column of $\mathbb V$ is a vector of ones and the $N$-dimensional $\mathbb V$ is a full-rank square matrix. Consequently, the transpose of $\mathbb V$ can be employed as an observation matrix by assigning an observation value $x_i$ to each microscopic state, with different observations corresponding to different powers of these observation values. Notably, the vector $\bm O$ contains the moments of \( x_i \) rather than \( x_i \) itself, indicating that the observation reflects the macroscopic properties. Consequently, $\mathbb{V}^\top{\bm{P}}=\bm{O}$ and $-\ln \bm{P} = \mathbb{V} \bm{B}$ can be derived, and the probability distribution of state $\sigma_i$ is given by 
\begin{equation}\label{Vdm}
    p_i = \frac{\exp(-\sum^{N}_{j=2} b_jx_i^{j-1} )}{\mathcal{Z}}.
\end{equation}
This provides a universal method for constructing observation matrices, which requires each microstate to have distinct observable values, with the observed averages corresponding to various orders of moments. For instance, the Vandermonde matrix naturally applies to particles undergoing random walks on a one-dimensional lattice with $N$ sites, where microstates are characterized by discrete positions such as $\{-2, -1, 0, 1, 2\}$ with a stable probability distribution. Consequently, the observed averages in $\bm{O}$ are the moments of particle positions $(1,\langle x \rangle,\langle x^2 \rangle,\langle x^3 \rangle,\langle x^4 \rangle)^\top$, from which the corresponding $\bm B$ can be derived to yield the associated Boltzmann distribution.

%\bibliographystyle{siam}
%\bibliographystyle{apsrev4-2}
%\raggedright
%\sloppy
\bibliography{main}

\end{document}